\pdfoutput=1
\documentclass[aps,prb,twocolumn,superscriptaddress,showpacs]{revtex4}

\usepackage{graphicx}

\begin{document}

\title{Oxide superlattices with alternating {\it p} and {\it n} interfaces}

\author{N. C. Bristowe}
\affiliation{Theory of Condensed Matter Group,
             Cavendish Laboratory, University of Cambridge, 
             JJ Thomson Ave, Cambridge CB3 0HE, UK}
\affiliation{Department of Earth Sciences, University of Cambridge, 
             Downing Street, Cambridge CB2 3EQ, UK}
\author{Emilio Artacho}
\affiliation{Department of Earth Sciences, University of Cambridge, 
             Downing Street, Cambridge CB2 3EQ, UK}
\author{P. B. Littlewood}
\affiliation{Theory of Condensed Matter Group,
             Cavendish Laboratory, University of Cambridge, 
             JJ Thomson Ave, Cambridge CB3 0HE, UK}

\date{\today}

\begin{abstract}
  The physics of oxide superlattices is considered for pristine (001) 
multilayers of the band insulators LaAlO$_{3}$ and SrTiO$_{3}$ with 
alternating $p$ and $n$ interfaces.
  A model of charged capacitor plates offers a simple paradigm to understand 
their dielectric properties and the insulator to metal transition (IMT) at 
interfaces with increasing layer thickness.
  The model is supported by first-principles results based on density-functional
theory.
  The charge at insulating interfaces is argued and found to be as predicted 
from the formal ionic charges, not populations.
  Different relative layer thicknesses produce a spontaneous polarization 
of the system, and allow manipulation of the interfacial electron gas.
  Large piezoresistance effects can be obtained from the sensitivity of the 
IMT to lateral strain.
  Carrier densities are found to be ideal for exciton condensation.
\end{abstract}

\pacs{73.21.Cd, 71.30.+h, 73.20.-r}

\maketitle

\section{INTRODUCTION}

  Recent technological advances have enabled the fabrication of high 
quality oxide multilayers, revealing a wealth of fascinating new physics.
  One striking example is the LaAlO$_{3}$/SrTiO$_{3}$ system (LAO/STO).
  In 2004, Ohtomo and Hwang~\cite{Ohtomo2004} discovered that the interface 
between these two perovskite band insulators can be conducting, depending on 
the termination of both materials.
  Many experimental and theoretical studies have followed 
\cite{Huijben2006,Nakagawa2006,Pauli2008,Reyren2007,Thiel2006,Albina2007,
Gemming2006,Janicka2009,Larson2008,Lee2008,Park2006,Pentcheva2006,
Pentcheva2008a,Pentcheva2009,Popovic2008, Chen2009}.
  Charge compensation at the interface of thick layers is required to avoid 
the so-called polar catastrophe.
  It arises from the diverging electrostatic potential caused by the
net electric charge at these interfaces resulting from the fact that the (001) 
monolayers of STO are neutral while the ones of LAO are charged.
  The TiO$_2$-LaO interface (termed $n$) needs 0.5 electrons per 
two-dimensional (2D) unit cell, and the SrO-AlO$_2$ interface ($p$) 
0.5 holes to avoid this polar catastrophe.
  Such numbers are based on formal ionic charges, i.e., 
Ti$^{+4}$, Sr$^{+2}$, La$^{+3}$, Al$^{+3}$, and O$^{-2}$,
although it is well known that the charge distribution in these materials 
is far from being so ideally ionic.
  Formal charges are often downscaled by so-called covalency parameters that 
aim to describe more realistic charges~\cite{Goniakowski2008}.

  These compensating electrons and holes are confined to the interface regions, but highly 
mobile in 2D, defining 2D electron and hole gases~\cite{Lee2008,Janicka2009}.
  Huijben \emph{et al}.~\cite{Huijben2006} (see \cite{Chen2009} for the 
theory) studied a system with one $p$ and one $n$ interface, finding 2D
conduction beyond a critical interface separation of five unit cells.
  Characteristics like carrier mobility depend on the carrier density,
which grows with separation beyond the critical value~\cite{Popovic2008}.
  Multilayers with both $p$ and $n$ interfaces with such a control 
of carrier density provide promising systems for obtaining the so-far 
elusive excitonic insulator~\cite{littlewood1996}.
  Recent experiments have demonstrated that oxygen vacancies can be controlled, while still allowing metallic conduction attributed to a 2D layer ~\cite{Thiel2006,Basletic2008}.



  This paper focuses on superlattices with alternating $p$ and $n$
interfaces.
  We consider pristine systems, with no point defects (the effect of 
oxygen vacancies~\cite{Park2006,Pentcheva2006} is discussed at the end). 
  Considering an equal thickness for both materials, an IMT is found by 
our calculations based on density-functional theory (DFT) for a 
thickness of just over eight unit cells.
  The electrostatic potential obtained in the calculations agrees
remarkably well with a simple model of capacitor
plates, giving an almost constant field of opposite sign in both materials,
which does not change with increasing thickness until the potential drop 
coincides with the band gap.
  At that critical thickness electrons transfer from the $p$ to the $n$ 
interface, making them conducting, and pinning the potential drop.

\section{METHOD}

  The DFT calculations were done using the local density 
approximation~\cite{Ceperley1980} and the {\sc Siesta} 
method~\cite{Ordejon1996,Soler2002}.
  Norm-conserving pseudopotentials~\cite{Troullier1991} were used, considering 
normal cores for O and Al, while semi-core electrons were included in the 
valence for La(5$s$5$p$), Sr(4$s$4$p$) and Ti(3$s$3$p$).
  Double-$\zeta$ polarized bases were used for valence 
electrons~\cite{Anglada2002}.
  Integrals in real space were performed on a mesh of 250 Ry
cutoff~\cite{Soler2002}, while Brillouin zone integrations were done on 
a $k$-mesh of 30 \AA\ cutoff~\cite{Moreno1992}.
  Four unit cells of STO in its ideal perovskite structure were layered on 
top of four of LAO, the 4/4 superlattice containing both TiO$_2$-LaO and 
SrO-AlO$_2$ terminations.
  The lateral lattice parameter was set to the theoretical average for
the two materials.
  The cell size was relaxed perpendicular to the interface ($z$) along
with the atomic positions until the forces were below 15 meV/\AA. 
  Samples for 8/8 and 12/12 unit cells were equally prepared.
  Other DFT studies ~\cite{Albina2007,Gemming2006,Janicka2009,
Larson2008,Lee2008,Park2006,Pentcheva2008a,Pentcheva2006,Popovic2008,
Pentcheva2009,Chen2009}, 
have focused on single interfaces of either kind, on arrays of repeated 
interfaces, or on alternating $p$ and $n$ interfaces.
  The latter~\cite{Park2006,Gemming2006,Albina2007} have considered
four unit cells or less, too thin to observe the physics discussed here.

\section{RESULTS AND DISCUSSION}

\subsection{Superlattice calculations}

  The band structures of the three superlattices are presented in Figure 1.
  The band gap is indirect and reduces with interface separation, closing
for the 12/12 system, with holes and electrons separated in reciprocal space.
  The behavior in real space is shown in Figure 2 for the 12/12 case
using the density of states projected onto each bilayer.
  The electrostatic potential is plotted alongside.

\begin{figure}[t]
\includegraphics[width=0.5\textwidth]{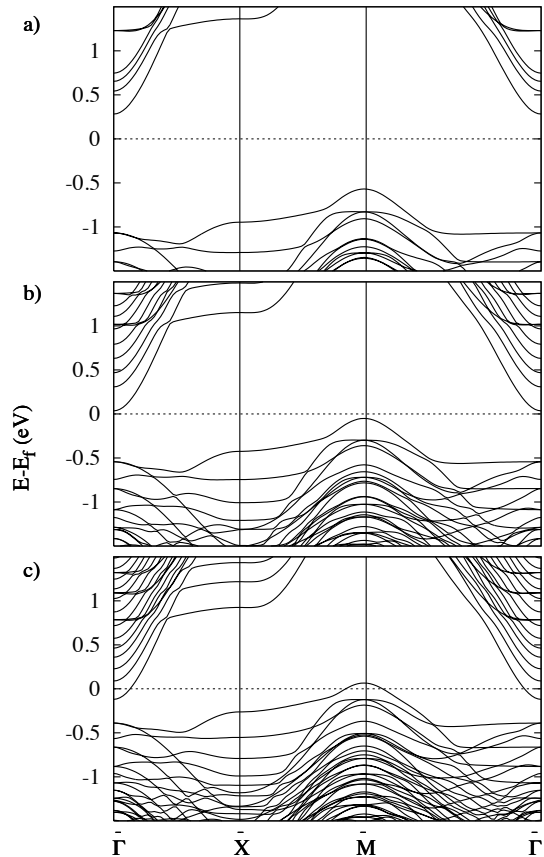}
\caption{\label{BANDS}{Band structures of three LaAlO$_3$/SrTiO$_3$
superlattices with (a) 4/4, (b) 8/8, and (c) 12/12 unit-cell thicknesses.}}
\end{figure}
 
\begin{figure*}[t]
\includegraphics[width=0.8\textwidth]{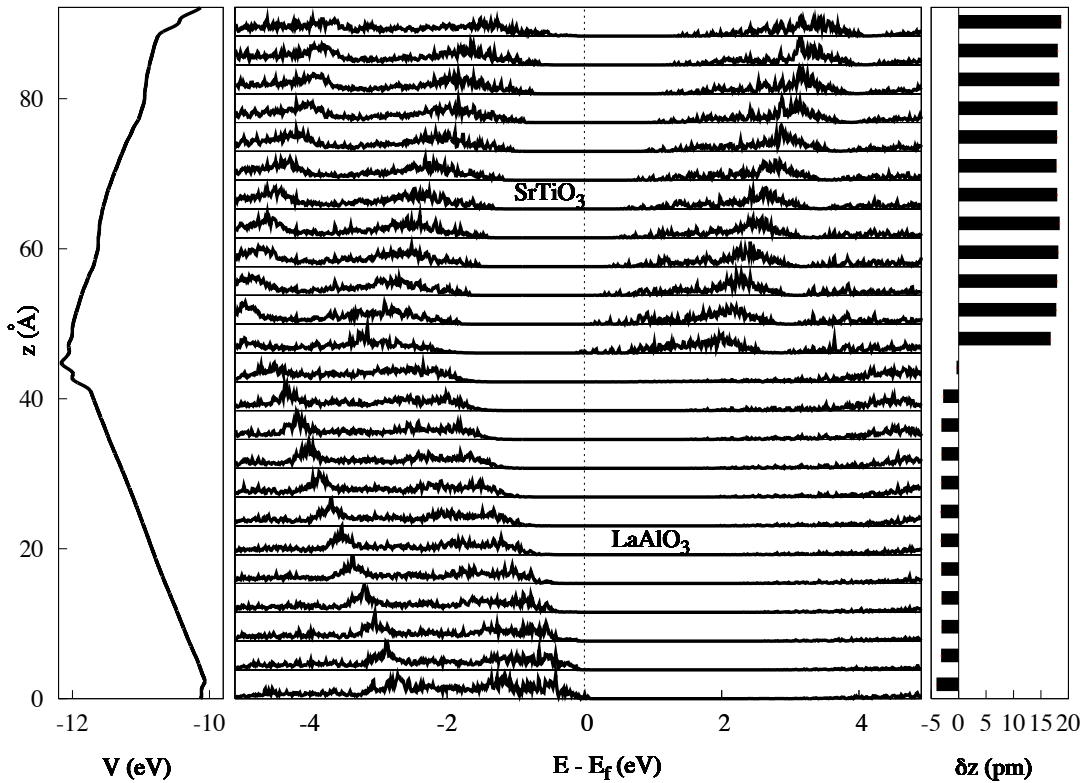}
\vspace{-0.5cm}
\caption{\label{DOS}{
  Centre: Density of states projected on unit-cell bilayers for the 
12/12 superlattice.
  Left: Averaged~\cite{BALDERESCHI1988} electrostatic potential energy for
electrons as a function of $z$.
  Right: The anion-cation splitting of the TiO$_2$ planes in SrTiO$_3$ 
and LaO planes in LaAlO$_3$ (larger splitting in each material), 
with $\delta z=z_{\rm cation}-z_{\rm anion}$.}}
\end{figure*}
  The physics of the problem is apparent in this figure.
  The net electric charge of chemical origin at the interfaces establish
electric fields among them, defining a zig-zag potential, which is closely
followed by the band structure of both materials.
  The IMT occurs when the amplitude of the zig-zag is larger than 
the band gap (the valence band offsets are small in this scale, 0.2 eV for the
$p$ and 0.0 eV for the $n$ interfaces).
  At this point holes appear confined in $z$ around the AlO$_2$ plane of
the $p$ interface, and around the ${\rm \bar M}$ point in the 2D
Brillouin zone, while electrons are confined to the TiO$_2$ plane at the 
$n$ interface, and around $\bar \Gamma$.
  The 2D electron gas (2DEG) is dominated by a Ti 3$d$ character. 
  It is nicely parabolic with an effective mass of 0.4 $m_e$.
  The several parallel sub-bands at this interface correspond to excitations 
under the effective confining potential for electrons along $z$ (Fig. 4a)
\footnote{Ref.~\onlinecite{Popovic2005} solves electron confinement in a 
V-shaped potential. Our 2DEG states are confined by different physics 
(Fig. 4a) and for a different system.}.

  Quantitative predictions for the IMT are biased by the known band-gap 
problem of Kohn-Sham fermions~\cite{Sham1983}.
  Constrained DFT calculations as used for charge-transfer 
systems~\cite{Wu2006} can be used for ours, but are beyond the scope of this 
work, its focus being the elucidation of the main mechanisms at play.
  In addition, it has been argued that, being LaTiO$_3$ a system with highly 
correlated electrons in the Ti 3$d$ band, methods addressing strong 
correlations are needed for our system.
  Note that this is not the case, however, since our IMT is in the limit 
of zero occupation of the 3$d$ band, hardly a correlated system, unlike
LaTiO$_3$, which has one 3$d$ electron per Ti atom.

  The deviations from the zig-zag behavior of the potential are small
in the scale of its amplitude, allowing the definition of a net electric 
field through each material.
  Its magnitude is nearly equal in both materials despite their different
dielectric response.
  Such response is illustrated in Figure 2 by the $z$-splitting of cations 
and anions at each layer, which is much larger for STO as expected.
  The fields obtained are 57.3 mV/\AA\ and 57.1 mV/\AA\ for the 
4/4 and 8/8 superlattices, while the 12/12 sees it 
reduced to 37.8 mV/\AA\ due to the partial charge back-transfer. 
  The field values correlate with the difference in sub-band
separation seen in Figure 1.
  An estimate of the 2DEG width $W$ is obtained from $W \sim
E_{ZPE} / {\cal E}$, i.e. the zero-point energy (ZPE) for the confining 
potential, over its slope, the electric field.
  Taking for 8/8, ${\cal E}=57$ mV/\AA\ and $E_{ZPE} \sim 0.2$ eV 
gives $W \sim 4$ \AA.

\subsection{Model of charged plates}
  
\begin{figure}[t]
\vspace{-5mm}
\includegraphics[width=0.5\textwidth]{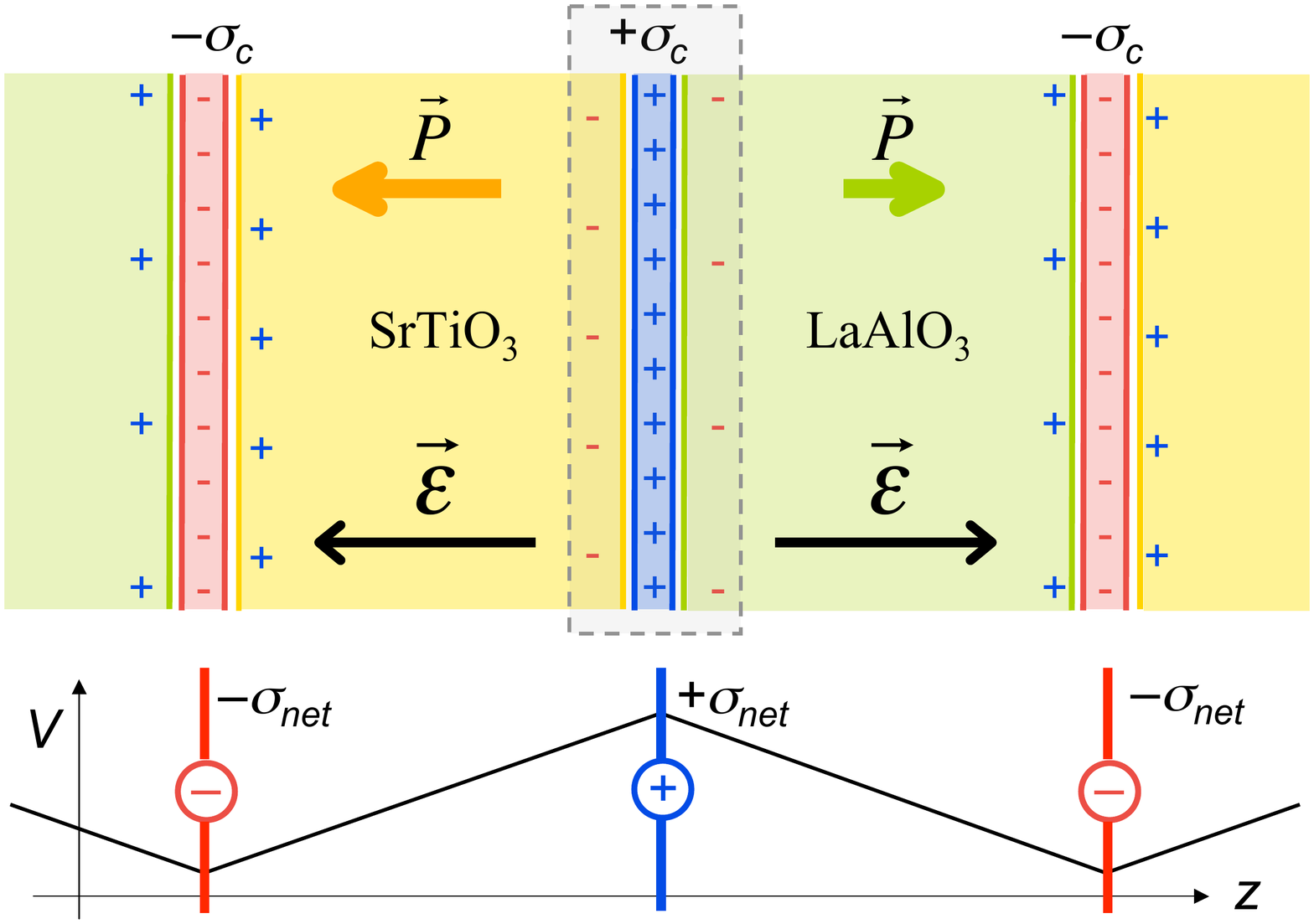}
\vspace{-8mm}
\caption{\label{PLATES}{(Color online) Capacitor-plates model of the
LAO/STO superlattice. 
  Above, the plates are indicated by the thinner bands, and
$\sigma_c$ indicates the charge of chemical origin attached to each, 
which is equal but of opposite sign for alternating interfaces.
  The box around the central plate indicates the surfaces for integration
of Gauss's law.
  The lower panel shows the net electrostatic potential $V$ for the system, 
which can be seen as arising from plates with
$\sigma_{\rm net}=\sigma_c - P_{\rm LAO}-P_{\rm STO}$, the latter $P_{\rm LAO}$
and $P_{\rm STO}$ referring to the magnitude of the respective polarizations
(notice that $V$ has opposite sign to what shown in the left panel of Fig.~2, 
which is $qV$). }}
\vspace{-3mm}
\end{figure}

  The physics described can be further analyzed with a simple model.
  The insulating system is modeled by a sequence of capacitor plates, 
one per interface, separated by dielectric material, as sketched in
Figure~\ref{PLATES}.
  Each plate has a planar charge density of chemical origin 
$\sigma_c$ (the $0.5e$ per interface unit cell described above), positive
at the $n$ interface, negative for the $p$.
  Using Gauss's law, periodic boundary conditions (PBC), and equal thicknesses,
there is a uniform electric field of magnitude ${\cal E}^0$ pointing outwards 
from the $n$ interface that satisfies
\begin{equation} 
\sigma_{c} - P_{\rm LAO} - P_{\rm STO}= 2\epsilon_0 {\cal E}^0 ,
\end{equation}
with $P_{\rm LAO}$ and $P_{\rm STO}$ the magnitude of the respective 
polarizations of both materials under the field, and $\epsilon_0$ the 
dielectric permittivity of vacuum. 
  The left hand side is what is indicated by $\sigma_{\rm net}$ in 
Figure~\ref{PLATES}.
  The behavior at the $p$ interface is exactly opposite,
with the field now pointing towards the plate.
  The fields have equal magnitude in both materials even if both
polarizations are different.

  Finding ${\cal E}^0$ requires knowing $P({\cal E})$ beyond linear
response, at least for STO.
  Under the strain conditions imposed by our geometry, the bulk of STO 
presents a spontaneous polarization along $z$~\cite{Haeni2004,Antons2005}, 
which allows a simpler modeling than for unstrained STO, and offers
an upper bound in the response, which corresponds to an upper bound to the 
critical thickness for the IMT.
  We take $P_{\rm LAO}=\epsilon_0 \chi_{\rm LAO} {\cal E}$ and 
$P_{\rm STO}= P_{\rm STO}^0 + \epsilon_0 \chi_{\rm STO} {\cal E}$.
  Defining $\kappa=2+\chi_{\rm STO}+\chi_{\rm LAO}$, equation 
1 becomes
\begin{equation}
{\cal E}^0=(\sigma_c-P_{\rm STO}^0) / \epsilon_0 \kappa.
\end{equation}

  We obtain $P_{\rm STO}^0= 0.309$ C/m$^2$ using the Berry phase 
approach~\cite{KINGSMITH1993} for bulk STO with the same strain conditions 
as in the multilayer.
  The lattice contribution to both susceptibilities is computed as in 
Ref.~\onlinecite{Antons2005} from the phonons and the Born effective charges,
which are obtained by finite differences of the forces and the 
polarization, respectively~\cite{Fernandez-Torre2004}. 
%
  The computed susceptibilities are $\chi_{33}^{ph}$(LAO)= 12.2 and 
$\chi_{33}^{ph}$(STO)= 24.7.
  To these we add the electronic contribution taking it from 
$\epsilon^{\infty}$(STO)= 5.18~\cite{ZHONG1994} and 
$\epsilon^{\infty}$(LAO)=4.77~\cite{Delugas2005}. 
  Using these bulk quantities and $\sigma_c=0.5e$ per interface
unit cell, an electric field of 57.4 mV/\AA\ is obtained, in 
excellent agreement with the superlattice DFT results.

\subsection{Chemical charge at interfaces}

  This agreement indicates that the physically meaningful charge $\sigma_c$ 
is the one predicted by formal ionic charges, rather than populations. 
  The reason is the same as for dopants in semiconductors, where
irrespective of the charge distribution around the dopant, the net 
charge is the result of counting core charges on one hand and electrons 
in the valence band on the other.
  Take the example of phosphorous as a donor in bulk silicon.
  Its +5 core is surrounded by the same set of bonds that surrounded the
Si +4 core it substitutes.
  The electron-pair clouds of these bonds are deformed (polarized), 
of course, but the number of electrons remains.
  The net effect is that the P donor creates an attractive potential for
electrons that corresponds to a charge of $+e$, not an effective charge of
any kind, the deformation of the bonds around being described by the 
dielectric constant of bulk silicon.

  These arguments were generalized to interfaces three decades 
ago~\cite{Baraff1977}.
  The valence-band electron counting was performed in terms of bands, but 
it can also be done using Lewis pairs or localized Wannier states.
  In LAO and STO the valence band corresponds to four Wannier states per 
oxygen, which gives exactly the same numbers as when using formal charges.
  The local charge neutrality picture~\cite{Goniakowski2008} can then
be used as the easiest way to see the interface charging, but using
formal charges, not effective ones.
  Whether atoms are chemically more or less ionic depends on
the shape and displacement of the Wanniers from the O atom towards 
the cations, but is irrelevant here.
  What is relevant is the number of electron pairs and the fact that they 
localize over lengths much smaller than the interface separation.

\subsection{Insulator to metal transition}

  The IMT occurs at the critical thickness of $d_c=\Delta/{\cal E}^0$,
where $\Delta$ is the gap (STO's), including the ZPE for
the confining potential at both interfaces (Fig. 4a).
  Taking the calculated ${\cal E}^0$, our DFT gap for strained 
STO (1.78eV), and both ZPEs as $\sim 0.2$ eV, a $d_c= 38$ \AA\ is obtained, 
$\sim 10$ unit cells.
  This value is sensitive to lateral strain, since $P({\cal E})$ 
depends on strain (especially for STO, so close to a ferroelectric
instability).
  Indeed, repeating the DFT calculation for 8/8 under the lateral lattice 
parameter of bulk STO gives conducting interfaces.
  This effect can be used to sense applied strain on a sample tuned to be 
close to the IMT (a piezoresistance device).

  Beyond the IMT, a constant density of states for holes and electrons is
assumed, modeling 2D parabolic bands.
  Considering electron transfer from the $p$ to the $n$ interface, 
equation (1) becomes 
\begin{equation}
\sigma_{c} -P_{\rm LAO}-P_{\rm STO}-({\cal E}d-\Delta)D= 2\epsilon_0 {\cal E}
\end{equation}
where $d$ is the interface separation, ${\cal E}$ the modified field, and
$D$ the density of states (taking equal effective masses for electrons 
and holes).
  Proceeding as for equation 2, 
\begin{equation}
{\cal E}=(\sigma_{c}+D\Delta-P_{\rm STO}^0)/(\epsilon_0 \kappa +D d).
\end{equation}
  The electric field vanishes as $1/d$ for large separations, the charge 
transferred tending to compensate the chemical charge.
  In our case, $D\Delta\gg\sigma_{c}$, and thus ${\cal E}/{\cal E}^0 
\sim d_c/d$, i.e. the voltage drop is essentially pinned by the gap.

\begin{figure}[t]
\vspace{-8mm}
\includegraphics[width=0.5\textwidth]{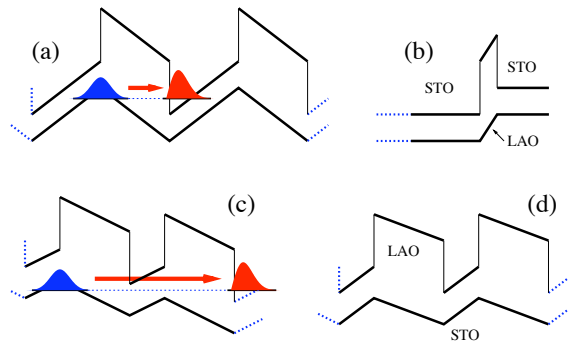}
\vspace{-20mm}
\caption{\label{MODEL}{(Color online) Model band gap versus $z$ for a 
superlattice with equal thicknesses (a), the thin film sample of 
ref.~\cite{Huijben2006} (b), and a superlattice with different thicknesses 
with open (c) and periodic (d) boundary conditions. 
  In (a) and (c) the arrow indicates the charge transfer leading to electron 
(red) and hole (blue) 2D gases.
  In both cases the thicknesses depicted give the onset of the charge 
transfer, corresponding to the insulator to metal transition. }}
\vspace{-3mm}
\end{figure}

\subsection{Extending the model to other systems}

  Consider now the model for other systems.
  Take a superlattice with thicker layers for LAO than for STO (Fig.~4c).
  It is analogous to a frozen $\langle 111 \rangle$ LO phonon in NaCl: the 
different separations of the charged planes gives a net electric field.
  PBC remove it as if putting the system between shorted
metal plates (Fig.~4d). 
  It is as a ferroelectric, except that switching the spontaneous polarization 
demands changing the thicknesses.
  Note that with PBC and thinner STO, ${\cal E}_{\rm STO} > {\cal E}_{\rm LAO}$.
  Without shorted plates the system is unstable to the appearance
of 2DEGs (Fig.~4c). 
 
  The $d_c$ found experimentally on the non-repeated system of 
ref.~\cite{Huijben2006} is much smaller than our superlattice result.
  They have one LAO layer interfacing $n$ to the STO substrate and $p$ to 
an STO overlayer (Fig.~4b).
  The field is zero in the substrate and overlayer, the $p$ and $n$ 
interfaces defining a capacitor only screened by LAO. 
  Thus, ${\cal E}=\sigma_c / \epsilon_0 (1 + \chi_{\rm LAO})$, more than twice 
as large as ${\cal E}^0$ of equation 2, giving the observed smaller $d_c$.
  A similar model has been very recently described for one single 
$n$ interface~\cite{Son2009}.


\section{FINAL REMARKS AND SUMMARY}

  The electron-hole interactions among 2D gases should establish excitons.
  Achievable exciton densities are favorable for exciton condensation,
the optimal density being~\cite{littlewood1996} $n_{opt}=(\pi a_0^2)^{-1}$,
with $a_0^2 \approx (a_B \epsilon / m^*)^2 + d^2$ (Bohr radius, dielectric
constant and effective
mass, respectively).
  For our system $n_{opt} \approx 0.002 e$/cell, well within range
($n_{12/12}=0.073 e$/cell).

  The IMT is affected by the presence of O vacancies.
Each donates two electrons to the $n$ interface giving rise to a 2DEG.
  Their appearance and location depends on their stability and
on kinetic effects like electromigration and sample-growth history. 
  Taking stability arguments only, the IMT via vacancies would happen 
when ${\cal E}^0 d \ge \mu_{\rm O}/2$, with ${\cal E}^0$ as in equation~2 and
$\mu_{\rm O}$ the formation energy of an O vacancy at the $p$ interface, 
which depends on the oxygen chemical potential at growth conditions. 
 
  In summary, the agreement between DFT and the model shows that it
contains the relevant physics of these superlattices, pointing to new 
science and applications by changing relative thicknesses, substrates, 
and stress. 

\begin{acknowledgments}
  We thank H. Y. Hwang, J. F. Scott, P. Ordejon, G. Catalan, N. D. Mathur, 
J. Junquera, W. E. Pickett and M. A. Carpenter for insightful discussions.
  We acknowledge support through EPSRC and the computing resources through 
the HPC facility in Cambridge.
\end{acknowledgments}


\end{document}